\documentclass[11pt,twoside]{article}  
\usepackage{asp2006}
\usepackage{adassconf}

\begin{document}   

%
%

\paperID{P2.12}

%

\title{Visualization of Complex Observational and Theoretical Datasets in
the Virtual Observatory}

%
%
%
%
%

\markboth{Chilingarian \& Zolotukhin}{Visualization of Complex Datasets in
the VO}

%
%
%
%

\author{Igor Chilingarian \altaffilmark{1,2}}
\author{Ivan Zolotukhin \altaffilmark{2}}

\altaffiltext{1}{Observatoire de Paris-Meudon, VO-Paris Data Centre; LERMA, UMR~8112, 61 Av. de l'Observatoire, Paris, 75014, France}
\altaffiltext{2}{Sternberg Astronomical Institute, Moscow State University, 13 Universitetsky prospect, Moscow, 119992, Russia}

%

\contact{Igor Chilingarian}
\email{Igor.Chilingarian@obspm.fr}

%
%
%

\paindex{Chilingarian, I.}
\aindex{Zolotukhin, I.}     

%



\begin{abstract}          
Our presentation is aimed at data centers providing access to complex
observational and theoretical data and to the users of these resources. We
show how to visualize complex datasets stored in the VO enabled data
archives using existing VO client software and PLASTIC, a prototype of an
application messaging protocol, for interaction between archive query
results and tools. We demonstrate how to display and explore observable IFU
datasets, provided within the ASPID-SR archive, using CDS Aladin, ESA
VOSpec, and VO-Paris Euro3D Client. In the second part of the paper
we show how to use TOPCAT for displaying results of N-body simulations of
galaxy mergers available in the HORIZON GalMer database. 
\end{abstract}

%
%

\section{Introduction}

At present the International Virtual Observatory has become a rapidly
growing initiative. Recently, several VO resources providing access to
complex observational and theoretical datasets have appeared. Providing the 
transparent and efficient data access and visualization mechanisms are
the crucial points for data sources to be used by the scientific community.

In this paper we demonstrate how to visualize complex observable and
theoretical datasets stored in the VO-enabled data archives using a
WEB-browser, existing VO client software and PLASTIC (PLatform for
Astronomical Tools Inter-Connection), a prototype of an application
messaging protocol, for interaction between archive query results and tools.
The technical details of the middle layer software implementation are given
in a paper ``Middleware for data visualization in VO-enabled data archives''
by Zolotukhin \& Chilingarian (this volume).

\section{ASPID-SR Archive}

ASPID-SR (Chilingarian et al. 2007a) is a prototype of an archive of
heterogeneous science ready data, containing observations obtained at the
Russian 6-m telescope. This resource provides the world largest collection
of science-ready 3D spectroscopic data, including about a hundred
integral-field unit (IFU) datasets mostly for extragalactic objects and
scanning Fabry-Perot interferometric observations (about 70 data-cubes for
nearby galaxies in H$\alpha$ and [OIII] emission lines).

ASPID-SR provides implementation for several existing IVOA standards:
Characterisation Data Model (Louys et al. 2007, one of the reference
implementations), Spectrum Data Model (McDowell et al. 2007) Simple Spectral
Access Protocol (Tody et al. 2007).

\begin{figure}[t]
\epsscale{0.7}
\plotone{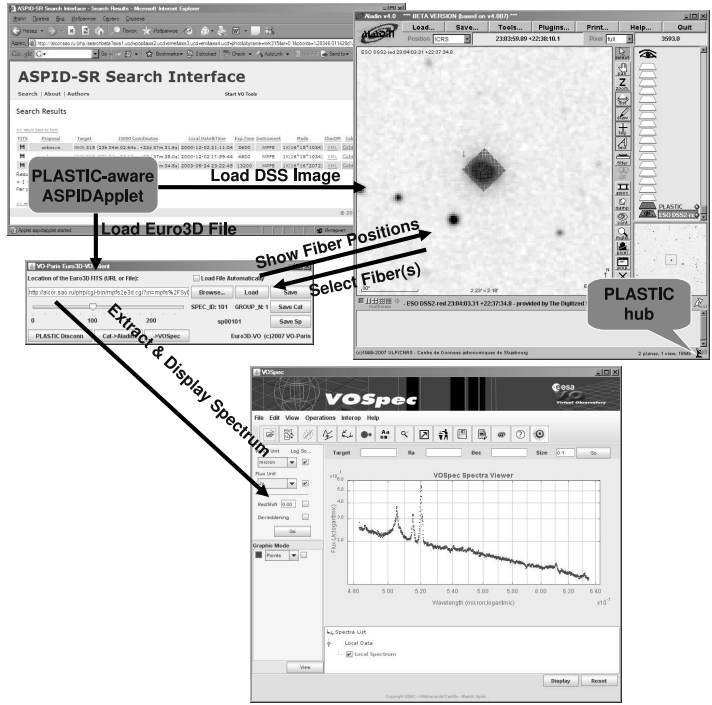}
\caption{Interaction between the ASPID-SR web-interface, CDS Aladin, ESA VOSpec,
and VO-Paris Euro3D Client.}
\label{P2.12_1}
\end{figure}

Interaction between VO client applications and the ASPID-SR archive
interface is implemented in several stages:
\begin{enumerate}
\item Querying the XML characterisation metadata (Zolotukhin et al. 2007) using the web interface.
\item Light-weight Java applet is integrated into the HTML pages, containing the
query response; it detects a PLASTIC hub, connects to it, and checks whether
other tools (Aladin, VOSpec, VO Paris Euro3D Client) are registered within
it. If the applications are not detected, they are started using
JavaScript and Java WebStart.
\item As soon as all the used applications have been started and registered
within the PLASTIC hub, a small script is sent to CDS Aladin to display the
DSS2 image of the area, corresponding to the position of the IFU
field of view. At the same time, the IFU dataset in the Euro3D FITS format 
(Kissler-Patig et al. 2004) is loaded into VO Paris Euro3D Client
(Chilingarian et al. 2007b).
\item Positions of IFU fibers are sent from VO Paris Euro3D Client to CDS
Aladin and overplotted on the DSS2 image.
\item User can interactively select either groups of fibers or individual ones
using CDS Aladin. An extracted spectrum (or co-added spectra of several
fibers) is sent to ESA VOSpec using PLASTIC by clicking on the corresponding
button in the user interface of VO Paris Euro3D Client.
\end{enumerate}

This implementation follows the principles of handling 3D spectral datasets,
proposed and described in Chilingarian et al. (2006)

\section{The Horizon GalMer Database}
The Horizon GalMer database (Di Matteo et al. 2007a) contains results of
N-body simulations of mergers of galaxies (Di Matteo et al. 2007b) of
different morphological types. To model the galaxy evolution, the Tree-SPH
code is used, where gravitational forces are calculated using a hierarchical
tree method and gas evolution is followed by means of smoothed particle
hydrodynamics. The first release of the data contains about 900 simulations
(with limited inclination angles of the orbits), in 50 to 70 snapshots each,
representing mergers of giant galaxies of different morphological types (E0
to Sd).

\begin{figure}[t]
\epsscale{0.8}
\plotone{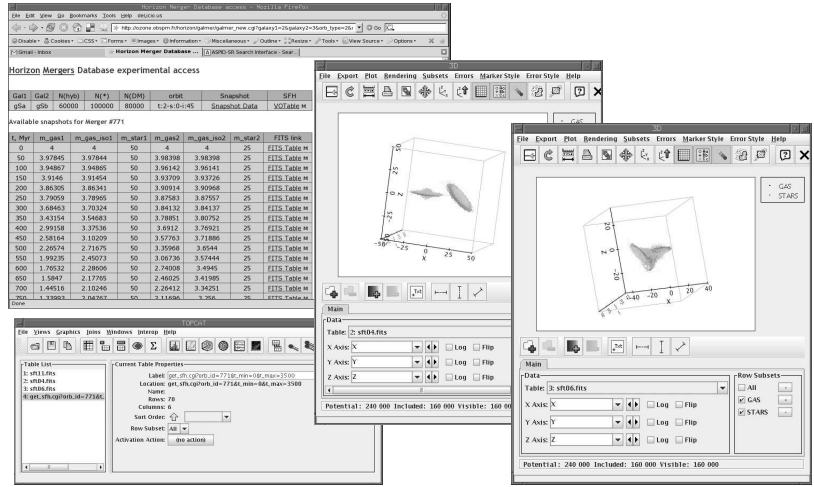}
\caption{This figure demonstrates interaction between the Horizon GalMer
Database, containing the results of N-body simulations, and TOPCAT used for
visualization. Two snapshots of a merger of giant Sa and Sc galaxies are
shown.} 
\label{P2.12_2}
\end{figure}

The web-based access to the simulation results is provided. The middleware
described in Zolotukhin \& Chilingarian (this conference) is used to provide
PLASTIC based communication between archive web-pages and TOPCAT used as a
tool for displaying the 3-dimensional snapshot datasets, as well as star
formation histories of the merging galaxies.

The Horizon GalMer database implements a prototype of IVOA Simple
Numeric Access Protocol Data Model (Lemson et al., in prep.) serialized as the
relational database schema.

\section{Summary}
The two implementations described above demonstrate that the observable and
theoretical datasets having complex structure can be discovered and accessed
at the present stage of the Virtual Observatory development, when not all
the interoperability standards are yet established. A WEB-browser and
existing client applications interacting via simple application messaging
protocol such as PLASTIC provide an infrastructure powerful enough for
scientific usage of the data sources in a VO framework.

\acknowledgments

Authors wish to thank ADASS organizing committee for the financial support
provided. Travel of IZ is also supported via RFBR grant 07-02-08846.

\end{document}